\documentclass{pasj00}
\draft

\begin{document}
\SetRunningHead{D. Yonetoku et al.}{Inefficient Electron Acceleration in GRB~060904A}
\Received{2000/12/31}
\Accepted{2001/01/01}

\title{Spectral evolution of GRB~060904A observed with 
{\it Swift} and {\it Suzaku}\\
--- Possibility of Inefficient Electron Acceleration}

\author{
Daisuke \textsc{Yonetoku},\altaffilmark{1,14}\email{yonetoku@astro.s.kanazawa-u.ac.jp (DY)}
Sachiko \textsc{Tanabe},\altaffilmark{1}
Toshio \textsc{Murakami},\altaffilmark{1}
Naomi \textsc{Emura},\altaffilmark{1}
Yuka \textsc{Aoyama},\altaffilmark{1}
Takashi \textsc{Kidamura},\altaffilmark{1}
Hironobu \textsc{Kodaira},\altaffilmark{1}
Yoshiki \textsc{Kodama},\altaffilmark{1}
Ryota \textsc{Kozaka},\altaffilmark{1}
Takuro \textsc{Nashimoto},\altaffilmark{1}
Shinya \textsc{Okuno},\altaffilmark{1}
Satoshi \textsc{Yokota},\altaffilmark{1}
Satoru \textsc{Yoshinari},\altaffilmark{1}
Keiichi \textsc{Abe},\altaffilmark{2}
Kaori \textsc{Onda},\altaffilmark{2}
Makoto S. \textsc{Tashiro},\altaffilmark{2}
Yuji \textsc{Urata},\altaffilmark{2}
Yujin E. \textsc{Nakagawa},\altaffilmark{3}
Satoshi \textsc{Sugita},\altaffilmark{3}
Kazutaka \textsc{Yamaoka},\altaffilmark{3}
Atsumasa \textsc{Yoshida},\altaffilmark{3}
Takuto \textsc{Ishimura},\altaffilmark{4}
Nobuyuki \textsc{Kawai},\altaffilmark{4}
Takashi \textsc{Shimokawabe},\altaffilmark{4}
Kenzo \textsc{Kinugasa},\altaffilmark{5}
Takayoshi \textsc{Kohmura},\altaffilmark{6}
Kaori \textsc{Kubota},\altaffilmark{7}
Kei \textsc{Sugiyasu},\altaffilmark{7}
Yoshihiro \textsc{Ueda},\altaffilmark{7}
Kensuke \textsc{Masui},\altaffilmark{8}
Kazuhiro \textsc{Nakazawa},\altaffilmark{8}
Tadayuki \textsc{Takahashi},\altaffilmark{8}
Shouta \textsc{Maeno},\altaffilmark{9}
Eri \textsc{Sonoda},\altaffilmark{9}
Makoto \textsc{Yamauchi},\altaffilmark{9}
Makoto \textsc{Kuwahara},\altaffilmark{10,11}
Toru \textsc{Tamagawa},\altaffilmark{10,11}
Daisuke \textsc{Matsuura}, \altaffilmark{12}
Motoko \textsc{Suzuki},\altaffilmark{13}
Scott \textsc{Barthelmy},\altaffilmark{14}
Neil \textsc{Gehrels},\altaffilmark{14} and
John \textsc{Nousek}\altaffilmark{15}
}
\altaffiltext{1}{Department of Physics, Kanazawa University, Kakuma, Kanazawa, Ishikawa 920-1192, Japan}
\email{yonetoku@astro.s.kanazawa-u.ac.jp (DY)}
\altaffiltext{2}{Saitama University, Sakura, Saitama 338-8570, Japan}
\altaffiltext{3}{Aoyama Gakuin University, Sagamihara, Kanagawa 229-8558, Japan}
\altaffiltext{4}{Tokyo Institute of Technology, Ohokayama, Meguro, Tokyo 152-8551, Japan }
\altaffiltext{5}{Gunma Astronomical Observatory, Takayama, Gunma 377-0702, Japan}
\altaffiltext{6}{Kogakuin University, Hachioji, Tokyo 192-0015, Japan}
\altaffiltext{7}{Kyoto University, Sakyo-ku, Kyoto 606-8502, Japan}
\altaffiltext{8}{JAXA/Institute of Space and Astronautical Science, Sagamihara, Kanagawa 229-8510, Japan}
\altaffiltext{9}{University of Miyazaki, Gakuen-kibanadai, Miyazaki 889-2192, Japan}
\altaffiltext{10}{RIKEN, 2-1 Hirosawa, Wako, Saitama 351-0198, Japan}
\altaffiltext{11}{Department of Physics, Tokyo University of Science, 1-3 Kagurazaka,
Shinjyuku-ku, Tokyo 162-8601, Japan}
\altaffiltext{12}{Osaka University, Toyonaka, Osaka 560-0043, Japan}
\altaffiltext{13}{Tsukuba Space Center, 1-1, Sengen 2chome, Tsukuba-city,
Ibaraki 305-8505, Japan}
\altaffiltext{14}{NASA/Goddard Space Flight Center, Greenbelt, MD 20771, USA}
\altaffiltext{15}{Pennsylvania State University, University Park, PA 16802, USA}

%


\maketitle

\begin{abstract}
We observed an X-ray afterglow of GRB~060904A with the {\it Swift} 
and {\it Suzaku} satellites. We found rapid spectral softening 
during both the prompt tail phase and the decline phase of an X-ray flare
in the BAT and XRT data. 
The observed spectra were fit by power-law photon indices which 
rapidly changed from $\Gamma = 1.51^{+0.04}_{-0.03}$ 
to $\Gamma = 5.30^{+0.69}_{-0.59}$ within a few hundred seconds 
in the prompt tail. This is one of the steepest X-ray spectra ever observed, 
making it quite difficult to explain by simple electron acceleration 
and synchrotron radiation. Then, we applied an alternative spectral fitting 
using a broken power-law with exponential cutoff (BPEC) model.
It is valid to consider the situation that the cutoff energy is equivalent
to the synchrotron frequency of the maximum energy electrons in their energy 
distribution. Since the spectral cutoff appears in the soft X-ray band,
we conclude the electron acceleration has been inefficient in the internal 
shocks of GRB~060904A. These cutoff spectra suddenly disappeared at 
the transition time from the prompt tail phase to the shallow decay one. 
After that, typical afterglow spectra with the photon indices of 2.0 are 
continuously and preciously monitored by both XRT and Suzaku/XIS up to 
1~day since the burst trigger time.
We could successfully trace the temporal history of two characteristic break 
energies (peak energy and cutoff energy) and they show the time dependence of 
$\propto t^{-3} \sim t^{-4}$ while the following afterglow spectra are quite stable. 
This fact indicates that the emitting material of prompt tail is due to completely 
different dynamics from the shallow decay component. Therefore we conclude 
the emission sites of two distinct phenomena obviously differ from each other.
\end{abstract}

\section{Introduction}
\label{sec:introduction}
Recent {\it Swift} \citep{gehrels04} observations reveal varying 
behavior with the early X-ray afterglows of Gamma-ray bursts (GRBs). 
\citet{nousek06} summarized that their X-ray lightcurves can be 
classified in three basic phases; a very steep decay, 
a shallow decay and the classical power-law decay, respectively. 
Additionally, strong X-ray flares during the very steep decay 
and the shallow decay phases are also found in many early 
X-ray afterglows. 

Jet breaks, as seen in many optical afterglows, have also been 
observed in X-ray lightcurves \citep{panaitescu}, 
but \citet{sato06} reported that 
the X-ray afterglow of GRB~050416A lacks its own jet break more than 
$\sim$~100~days after the burst trigger time. This fact means that 
the jet opening half angle must be $\theta \ge 23$~degree, 
which is much wider than typical opening angles of GRB jets.
On the other hand, {\it Swift} and {\it Suzaku} 
combined observations of GRB~060105, \citet{tashiro07} found a very early jet break 
less than 0.04~day after the GRB trigger time. 
Therefore we must study these time profiles on a case by case basis. 
We need much more information about the early X-ray 
afterglows to comprehend these complex characteristics.

Several authors reported significant spectral softening 
during the early X-ray afterglows. Especially, \citet{bbzhang06} 
performed simultaneous spectral analyses for the brightest 
17~cases of X-ray afterglows observed by the {\it Swift}/XRT. 
In their report, 10 of 17 samples show strong spectral evolution 
while the others have no evolution. These spectral evolutions 
are generally explained by a model based on the curvature effect of emitting 
region \citep{fenimore96, kumar00, dermer04, yamazaki06}. 
This model implies that the temporal index ($\alpha$) and the spectral energy 
index ($\beta$) have to show co-evolution as $\alpha = \beta + 2$ 
for the simplest curvature effect \citep{liang06}. 
However, the observed spectral evolution does not match with the 
property of temporal decline, so \citet{bbzhang06} introduced 
another hidden component - the central engine afterglow which 
may be related to the continuous activity of GRB central engines.
These characteristics must be investigated in detail because 
the cases with spectral softening make up a large fraction of 
long GRBs.

Using BATSE data, \citet{ryde05} investigated the possibility 
of the existence of thermal emission in the prompt GRB spectra. 
He proposed a prompt emission spectrum composed of a thermal (blackbody) 
spectrum combined with a non-thermal (power-law) one. 
\citet{butler06} also explained the anomalous soft X-ray spectrum 
of GRB~060218 as the thermal plus non-thermal model. He suggested that 
the X-ray afterglow is dominated by thermal emission with an effective 
temperature of $kT \sim 0.3~{\rm keV}$. 
Even after introducing the thermal blackbody model, 
the spectral evolution of the power-law component still remains 
and the photon indices change from $\Gamma = 1.5$ to $\Gamma = 3.4$. 
Therefore, the thermal model is not sufficient to explain all the spectral softening 
and additional spectral evolution is required. 

In this paper, we show observation results on GRB~060904A 
with the {\it Swift} and {\it Suzaku} satellites. 
The spectral photon index achieves $\Gamma = 5.30^{+0.69}_{-0.59}$
which is one of the strongest spectral softenings ever observed.
In the next section, we summarize X-ray observations with
the {\it Swift} and {\it Suzaku} satellites for this event. 
In the third section, we show detailed data reductions for both
datasets. 
We fit two individual models (the single power-law, and 
a broken power-law with exponential cutoff: BPEC) to the observed 
spectra. The fitting results and spectral properties are summarized
in sections 4 and 5. We succeeded in separating the X-ray afterglow 
component from the prompt tail and/or X-ray flare. In the 6th section,
we show an interesting spectrum during the time when the X-ray afterglow and 
prompt tail emission co-exist. In the final session, we discuss  
the observed spectral and temporal properties of GRB~060904A,  
and we suggest the possible presence of a maximum energy cut-off in 
the electron energy distribution. 

\section{Observations}
\label{sec:observation}
GRB~060904A was detected by the BAT instrument aboard the {\it Swift} 
satellite at 2006/09/04 01:03:21 (UT), and localized at 
${\rm R.A.} = 15^h 50^m 58^s,~{\rm Dec.} = +44^d 57' 57"$ (J2000) 
with an uncertainty of 3~arcminutes. The BAT lightcurve shows several 
little peaks and then an intense flare composed of multiple spikes at 
$t - t_{trigger} \sim 55~{\rm sec}$. Here $t_{trigger}$ is the GRB 
trigger time. The burst duration time measured in the BAT energy 
range was about $T_{90} \sim 85~{\rm sec}$,
where $T_{90}$ is measured as the duration of the time interval 
during which 90~\% of the total observed counts have been detected.

The XRT automatically started a follow-up observation from 
$t - t_{trigger} = 66~{\rm sec}$ in windowed timing mode. 
A bright X-ray afterglow was found at ${\rm R.A.} = 15^h 50^m 54^s.9, 
{\rm Dec.} = +44^d 59' 07''.8$ (J2000) with an estimated uncertainty 
of 5.4 arcseconds. The initial flux measured by the first 0.1~sec image 
was $2.7 \times 10^{-8}~{\rm erg~cm^{-2}s^{-1}}$ (0.2--10 keV band). 
Unfortunately, {\it Swift} entered the South Atlantic Anomaly at
01:43:52 (UT), and the XRT observation was interrupted by the next 
GRB~060904B triggered at 02:31:03 (UT). Therefore XRT data lasted 
only about 2000~seconds after the burst trigger time.

\begin{figure}
\rotatebox{0}{\includegraphics[width=100mm]{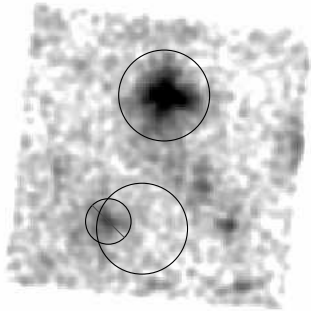}}
\caption{The image of X-ray afterglow of GRB~060904A observed
with the {\it Suzaku}/XIS. The brightest source in the frame 
is the X-ray afterglow, and we extract X-ray events within 
2.89~arcmin radius as source signals. The background region is 
determined at the opposite side across the optical axis of 
the X-ray telescope. Since we found a dim source within the 
background region, it was removed as shown by a small circle 
with a diagonal line.}
\label{fig:xis-image}
\end{figure}

The X-ray intensity reported by the {\it Swift}/XRT team was quite 
bright and also the GRB trigger time satisfied  criteria 
for a target of opportunity (ToO) observation by the Japanese X-ray 
satellite {\it Suzaku}. {\it Suzaku} began the follow-up observation 
using an HXD nominal pointing mode from 
10:29:46~(UT) to 2006/09/05 05:03:46 (UT) with net exposure time 
of 30.4 ksec. We confirmed the X-ray afterglow found by 
the {\it Swift}/XRT within the XIS field of view as shown in
figure~\ref{fig:xis-image}.
Therefore {\it Suzaku} covered the late time behavior 
of the X-ray afterglow which could not be observed by {\it Swift}.

No optical counterpart was reported, and this burst was categorized 
as an optically dark GRB. So we have no direct information about the 
redshift. The {\it Subaru} telescope observed the field of GRB~060904A 
with the MOIRCS near-infrared camera $\sim 5$ hours after 
the burst trigger time. An extended object was found within 
the XRT error circle as reported by \citet{aoki06}, 
but it is still in debate whether 
the object is a host galaxy of GRB~060904A or not. 
Konus-Wind measured the spectral peak energy as 
$E_{p} = 163 \pm 31~{\rm keV}$ in the 20~keV -- 2~MeV energy range 
\citep{golenetskii06}. The Wide-band All-sky Monitor (WAM) aboard 
{\it Suzaku} also detected the prompt emission \citep{tashiro06}. 
Using luminosity indicators (e.g. Amati and Yonetoku relation), 
a possible redshift is estimated as pseudo-$z = 1.84 \pm 0.85$ 
\citep{pelangeon06}. Hereafter we will assume this redshift when 
we discuss physical quantities measured in the rest frame of the GRB.

\begin{figure}
\rotatebox{270}{\includegraphics[width=65mm]{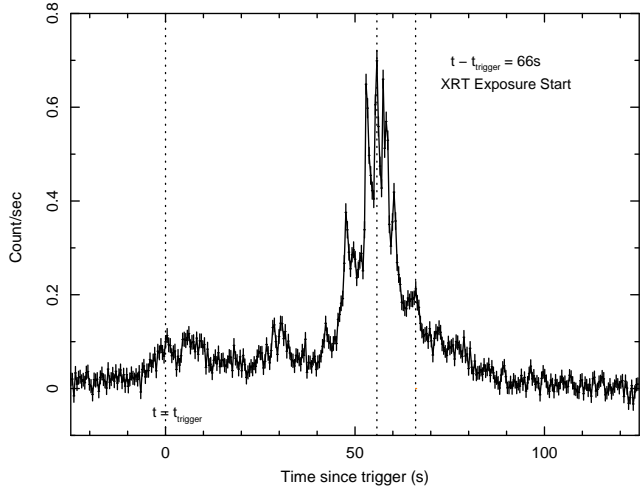}}
\caption{The BAT lightcurve of prompt emission of GRB~060904A.
The XRT follow-up observation was started at $t-t_{trigger} = 66$~sec
during the prompt tail.}
\label{fig:bat-lc}
\end{figure}

\section{Data Reduction}
\label{sec:reduction}
The {\it Swift}/BAT and XRT data were analyzed by standard analysis 
tools within the heasoft~6.2.0 packages
\footnote{http://swift.gsfc.nasa.gov/docs/swift/analysis/}.
For the windowed timing mode data of XRT, using xselect software, 
we extracted both X-ray afterglow and background spectra within 
rectangular regions of 40~pixels in width with a height large enough 
to include all the photons.
We used the energy response matrix file (RMF) released by 
the XRT team, and the ancillary response file (ARF) was 
generated for each spectral file with the ``xrtmkarf'' command 
including the latest calibration database.

\begin{figure}
\rotatebox{0}{\includegraphics[width=120mm]{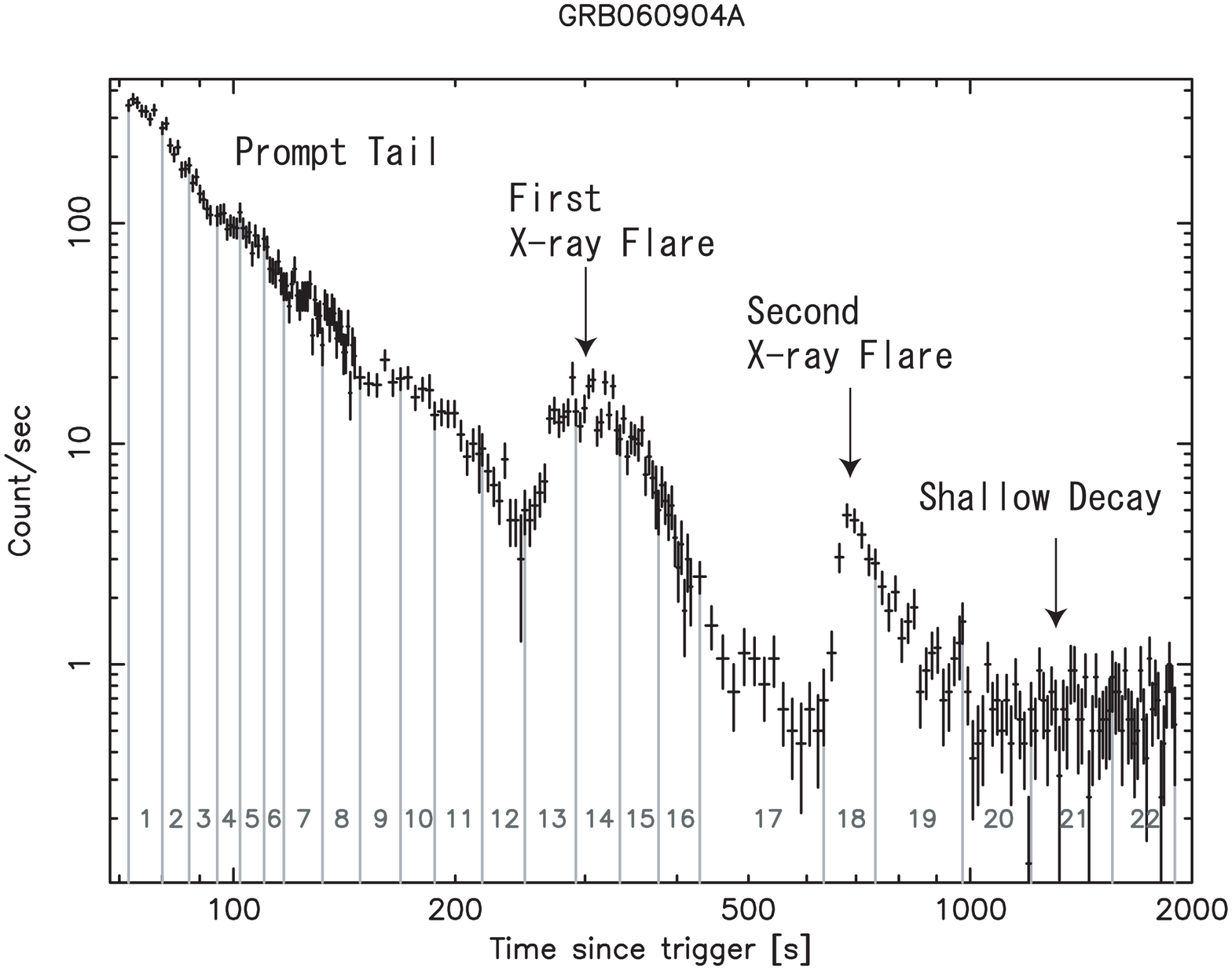}}
\caption{The lightcurve of the early X-ray afterglow of GRB~060904A 
observed with {\it Swift}/XRT (0.5--10 keV range). We performed time 
resolved spectral analyses for 22 divisions as shown in this figure.}
\label{fig:xrt-lc}
\end{figure}

We divided the entire XRT data into 22 time intervals as shown 
in figure~\ref{fig:xrt-lc}. Since the X-ray count rate is quite high 
just after the start of XRT's pointing observation, we removed 
the brightest 4~pixels from the central part of the source 
region to deal with pile-up effects 
until $t - t_{trigger} \le 86.55~{\rm sec}$. 
As shown in the X-ray lightcurve, we can recognize 2 remarkable X-ray 
flares around 300~sec and 700~sec. We divided the XRT data to trace 
the arising and decaying behaviors for these two X-ray flares.
Although the first steep decay (prompt tail) has several fluctuations 
in X-ray intensities, we will not discuss their temporal behavior 
in detail because their emission power changes are dominated by the baseline 
variation in the main prompt tail. 

We used data resulting from processing version rev1.2
for both the {\it Suzaku}/XIS \citep{koyama07} and HXD \citep{takahashi07}, 
and we also used the heasoft~6.2.0 packages when we performed the data 
reduction\footnote{http://www.astro.isas.jaxa.jp/suzaku/analysis/}.
For the XIS data, as shown in figure~\ref{fig:xis-image}
we took the source region as a circle of 
2.89~arcmin in radius, and the background 
region as the same radius from the opposite side across the optical 
axis of the X-ray telescope. Since we found a dim source in the 
background region, we removed it with smaller circle of about 2~arcmin 
in radius. We generated an RMF and an ARF with ``xisrmfgen'' 
and ``xissimarfgen'' tools including the latest calibration 
database, respectively. For the XIS data, we divided it 
into 5 time intervals and performed spectral analyses.

The 16 HXD-PINs in the well unit number of W0 are operated with bias 
voltage of 400~V since 2006 May 24, to suppress the rapid increase 
of noise events possibly caused by in-orbit radiation damage.
The others were operated with a nominal bias of 500~V. 
Therefore we excluded the data obtained by 16 PINs in the W0 unit 
to avoid a large uncertainty in in-orbit calibrations.
We used the non X-ray background (NXB) event data produced by 
the HXD team, and performed standard data reduction for remaining
48~PINs in W1, 2, and 3 units. The cosmic X-ray background (CXB) 
is not included in the NXB event data, so we modeled the functional
form of the CXB based on past observations.
The recommended RMF for a point source observed in the HXD nominal 
mode was adopted when we performed spectral analyses.
Spectral fitting was performed using the XSPEC~12 packages 
for both {\it Swift} and {\it Suzaku} data. 
Hereafter the quoted errors are at the 90~\% confidence level.

\section{Spectral and Temporal Analyses}
\label{sec:analyses}

\subsection{Single Power-law Fitting}
We investigated spectral evolution for these 27 datasets with 
a simple absorbed power-law model. Hereafter, we assume galactic 
absorption $N_{\rm H}^{gal} = 1.41 \times 10^{20}~{\rm cm^{-2}}$ as 
fixed parameters \citep{dickey90}, and the extra galactic 
absorption $N_{\rm H}^{ext}$ as a free parameter.
We also assume a possible redshift of pseudo-$z = 1.84$ 
by \citet{pelangeon06}. We performed spectral fitting in 0.5--10~keV 
for both the XRT and XIS.

\begin{figure}
\rotatebox{270}{\includegraphics[width=120mm]{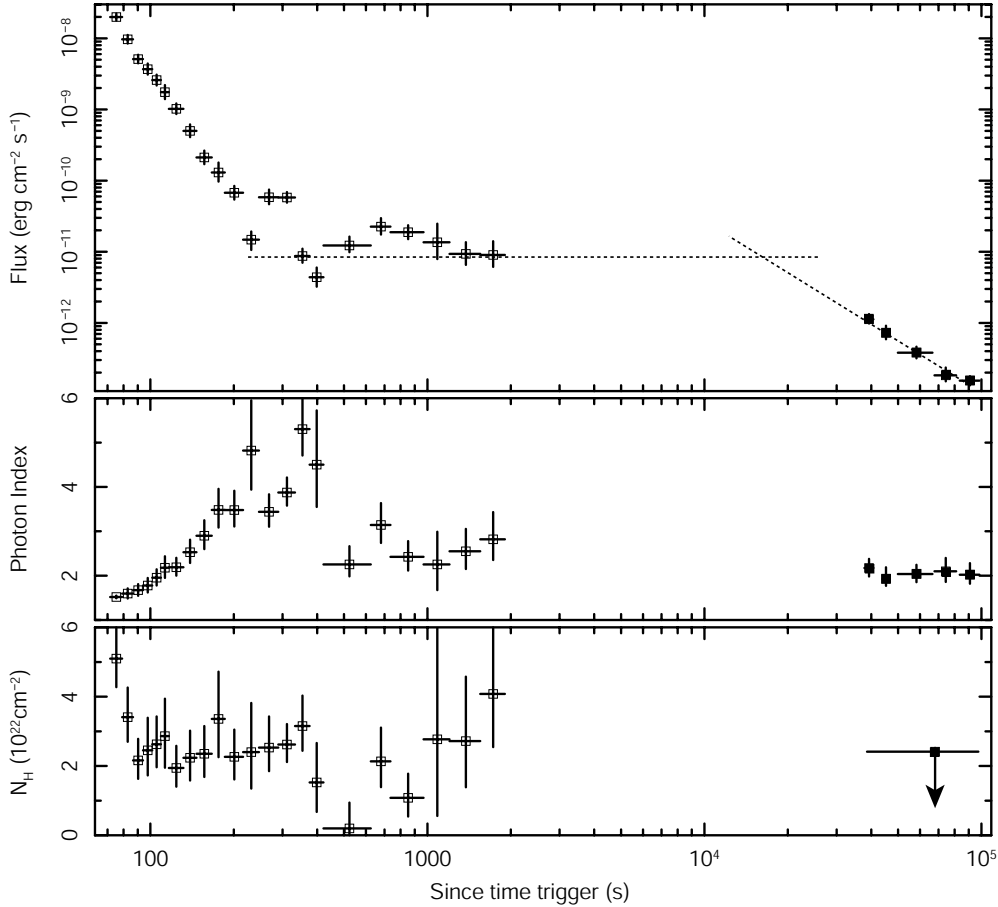}}
\caption{The temporal histories of 2--10~keV energy flux (top), 
photon index (middle) and the extragalactic absorption (bottom) 
for the X-ray afterglow of GRB~060904A, respectively.
The open and filled squares indicate the {\it Swift}/XRT 
and the {\it Suzaku}/XIS observations.}
\label{fig:single-power}
\end{figure}

In figure~\ref{fig:single-power}, we show the temporal 
histories of the X-ray energy flux in 2--10~keV band, 
the photon index ($\Gamma$), and the extragalactic absorption 
($N_{\rm H}^{ext}$), 
respectively. The temporal index of the initial steep decay phase can 
be described by single power-law as $(t - t_{trigger})^{-6.0 \pm 0.2}$ 
for the time interval No.~1--12. 
We can see that the flux level in the shallow decay phase is almost 
constant. For the late time {\it Suzaku} observation, we can describe 
the temporal decline as $(t - t_{trigger})^{-2.4 \pm 0.3}$, which is 
equivalent to the steeper decay phase after the jet break. 
Combined with {\it Swift} and {\it Suzaku} data, we can estimate
a possible jet break time as 
$1.5 \times 10^{4} < t_{j} < 3.8 \times 10^{4}~{\rm sec}$ 
($0.177 < t_{j} < 0.440~{\rm day}$). 

We can obtain acceptable fitting results for the single power-law model.
However the photon indices show a remarkable and continuous softening 
during the prompt tail phase from $\Gamma = 1.5$ to $\Gamma = 4.8$. 
Starting with the first X-ray flare, the spectrum slightly 
became harder, and showed re-softening during its decaying phase.
The photon index achieved $\Gamma = 5.30^{+0.69}_{-0.59}$ which is 
one of the steepest case ever observed. 
After the second X-ray flare, the photon indices are rapidly 
settled down to $\Gamma \sim 2$, which is a typical value for 
most X-ray afterglows, including the late time afterglow
observed with {\it Suzaku}.

The column density $N_{\rm H}^{ext}$ stays almost constant
during the XRT observation, but we found the existence of some 
discrepancies between the XRT and XIS. 
A contamination of carbon and oxygen on the optical blocking filter
of each XIS must be appropriately treated because it affects
the absorption quantity in the low energy band. Although we include 
the contamination effect using the ``xissimarfgen'' tool with 
the latest calibration files, a systematic uncertainty exists 
about equivalent hydrogen column density of 
$6 \times 10^{20}~{\rm cm^{-2}}$ in the observer's rest frame.
For a high redshift object, the systematic uncertainty of 
contamination should be converted by a factor of $(1+z)^{3}$.
Therefore the estimated column density $N_{\rm H}^{ext}$ is 
influenced by a large systematic error of about 
$\pm 1.4 \times 10^{22}~{\rm cm^{-2}}$.
As a result, we can only set an upper-limit of 
$N_{\rm H}^{ext} < 2.3 \times 10^{22}~{\rm cm^{-2}}$
for the {\it Suzaku} observation.
The best fit parameters are summarized in table~\ref{table:single-po}.


\begin{table}
\caption{Fitting results with the single power-law.\label{table:single-po}}
\begin{center}
\begin{tabular}{lcccccc}
\hline\hline
No.(Detector) & $t-t_{trigger}$ & $\Delta t$ & Flux (2--10 keV) & $\Gamma$ & $N_{\rm H}^{\rm ext}$ & $\chi^{2}_{\nu}$/dof\\
    & (sec) & (sec) & (${\rm erg/cm^2/s}$) & & ($10^{22}~{\rm cm^{-2}}$) &\\
\hline
Prompt Tail ({\it Swift}) &&&&& & \\
1 (XRT+BAT) & 75.3  & 7.5 &  $(1.99^{+0.24}_{-0.21}) \times 10^{-8}  $&  $1.51^{+0.04}_{-0.03}$ & $5.10^{+0.99}_{-0.82}$ & 1.00/142\\
2 (XRT+BAT) & 82.8  & 7.5 &  $(9.67^{+1.49}_{-1.23}) \times 10^{-9}  $&  $1.59^{+0.11}_{-0.11}$ & $3.41^{+0.85}_{-0.71}$ & 1.12/112\\
3 (XRT+BAT) & 90.3  & 7.5 &  $(5.11^{+0.77}_{-0.66}) \times 10^{-9}  $&  $1.67^{+0.13}_{-0.12}$ & $2.16^{+0.62}_{-0.53}$ & 1.10/101\\
4 (XRT+BAT) & 97.8  & 7.5 &  $(3.68^{+0.75}_{-0.59}) \times 10^{-9}  $&  $1.77^{+0.17}_{-0.15}$ & $2.45^{+0.93}_{-0.72}$ & 0.89/69\\
5 (XRT) & 105.3 & 7.5 &  $(2.58^{+0.50}_{-0.42}) \times 10^{-9}  $&  $1.95^{+0.18}_{-0.17}$ & $2.62^{+0.80}_{-0.66}$ & 1.15/34\\
6 (XRT) & 112.8 & 7.5 &  $(1.75^{+0.46}_{-0.35}) \times 10^{-9}  $&  $2.17^{+0.25}_{-0.23}$ & $2.86^{+1.08}_{-0.91}$ & 0.76/27\\
7 (XRT) & 124.1 & 15.0 & $(1.02^{+0.19}_{-0.16}) \times 10^{-9}  $&  $2.18^{+0.21}_{-0.19}$ & $1.94^{+0.63}_{-0.54}$ & 1.14/39\\
8 (XRT) & 139.1 & 15.0 & $(4.99^{+1.16}_{-0.92}) \times 10^{-10} $&  $2.52^{+0.28}_{-0.24}$ & $2.23^{+0.78}_{-0.65}$ & 1.21/26\\
9 (XRT) & 156.6 & 20.0 & $(2.11^{+0.53}_{-0.41}) \times 10^{-10} $&  $2.89^{+0.35}_{-0.30}$ & $2.35^{+0.80}_{-0.66}$ & 1.06/21\\
10 (XRT)& 176.1 & 19.0 & $(1.30^{+0.49}_{-0.34}) \times 10^{-10} $&  $3.48^{+0.47}_{-0.39}$ & $3.35^{+1.36}_{-1.10}$ & 0.95/18\\
11 (XRT)& 200.8 & 30.4 & $(6.74^{+1.70}_{-1.33}) \times 10^{-11} $&  $3.47^{+0.43}_{-0.37}$ & $2.26^{+0.78}_{-0.65}$ & 0.72/19\\
12 (XRT)& 231.3 & 30.4 & $(7.52^{+3.44}_{-2.18}) \times 10^{-12} $&  $4.82^{+1.13}_{-0.88}$ & $2.40^{+1.41}_{-1.05}$ & 0.87/14\\
\hline
First Flare ({\it Swift})&&&&&&\\
13 (XRT)& 268.1 & 43.0 & $(5.84^{+1.64}_{-1.17}) \times 10^{-11} $&  $3.43^{+0.39}_{-0.33}$ & $2.53^{+0.89}_{-0.68}$ & 1.25/22\\
14 (XRT)& 311.1 & 43.0 & $(5.79^{+1.11}_{-0.90}) \times 10^{-11} $&  $3.87^{+0.33}_{-0.29}$ & $2.61^{+0.59}_{-0.50}$ & 1.10/34\\
15 (XRT)& 354.1 & 43.0 & $(8.71^{+2.32}_{-1.72}) \times 10^{-12} $&  $5.30^{+0.69}_{-0.59}$ & $3.15^{+0.87}_{-0.71}$ & 1.26/21\\
16 (XRT)& 398.6 & 46.0 & $(4.38^{+1.61}_{-1.15}) \times 10^{-12} $&  $4.50^{+1.22}_{-0.95}$ & $1.52^{+1.13}_{-0.85}$ & 1.69/8\\
17 (XRT)& 522.9 & 202  & $(1.22^{+0.41}_{-0.24}) \times 10^{-11} $&  $2.25^{+0.41}_{-0.27}$ & $0.19^{+0.74}_{-0.19}$ & 1.19/10\\
\hline
Shallow Decay ({\it Swift})&&&&&&\\
Including Second Flare &&&&&&\\
18 (XRT)& 679.8 & 111  & $(2.25^{+0.72}_{-0.51}) \times 10^{-11} $&  $3.14^{+0.49}_{-0.40}$ & $2.13^{+0.97}_{-0.74}$ & 0.84/17\\
19 (XRT)& 851.8 & 234  & $(1.88^{+0.48}_{-0.38}) \times 10^{-11} $&  $2.42^{+0.35}_{-0.31}$ & $1.08^{+0.69}_{-0.54}$ & 1.00/17\\
20 (XRT)& 1085  & 234  & $(1.36^{+1.12}_{-0.57}) \times 10^{-11} $&  $2.25^{+0.73}_{-0.58}$ & $2.76^{+3.21}_{-2.21}$ & 0.40/6\\
21 (XRT)& 1377  & 350  & $(9.33^{+4.31}_{-2.80}) \times 10^{-12} $&  $2.55^{+0.50}_{-0.40}$ & $2.71^{+1.86}_{-1.33}$ & 1.03/12\\
22 (XRT)& 1728  & 352  & $(9.00^{+5.06}_{-2.88}) \times 10^{-12} $&  $2.81^{+0.61}_{-0.46}$ & $4.08^{+2.19}_{-1.53}$ & 0.92/13\\
\hline
Late Afterglow ({\it Suzaku}) &&&&& & \\
23 (XIS)& 39262 & 2942 & $(1.13^{+0.20}_{-0.15}) \times 10^{-12}$ &  $2.17^{+0.21}_{-0.19}$ &  & 1.06/39\\
24 (XIS)& 45210 & 2551 & $(7.24^{+1.89}_{-1.36}) \times 10^{-13}$ &  $1.93^{+0.27}_{-0.15}$ &  & 0.90/35\\
25 (XIS)& 58234 & 16631& $(3.81^{+0.82}_{-0.65}) \times 10^{-13}$ &  $2.04^{+0.21}_{-0.18}$ & $< 2.30$ & 0.99/38\\
26 (XIS)& 74320 & 13365& $(1.83^{+0.53}_{-0.33}) \times 10^{-13}$ &  $2.10^{+0.30}_{-0.24}$ & & 1.15/32\\
27 (XIS)& 90868 & 14787& $(1.54^{+0.24}_{-0.22}) \times 10^{-13}$ &  $2.02^{+0.26}_{-0.20}$ & & 1.11/31\\
\hline
\end{tabular}
\end{center}
$^{*}$ this value was measured from the averaged XIS spectrum.
Because of the contamination effect (see text), this value has a large
systematic uncertainty.
\end{table}

\subsection{Average Spectrum Obtained by {\it Suzaku}/XIS and HXD}
In the previous subsection, for the {\it Suzaku} data, we divided 
the entire spectra into 5~time intervals as listed in 
table~\ref{table:single-po}. We successfully fitted these spectra 
with the absorbed power-law model, and found no significant 
spectral evolution. Therefore, we investigated the average spectrum
for the late time X-ray afterglow to determine the important spectral parameters.

We also performed spectral fitting for the time averaged data from 
both the {\it Suzaku}/XIS (0.5--10.0~keV) and HXD-PIN (12--60~keV). 
For the HXD-PIN fitting, we subtract only the NXB from the observed data, 
and the CXB is included as the fixed function. A reproduction for the 
NXB modeling is 
about 4~\% for the nominal HXD observations, so we include this 
systematics when we subtract the NXB spectrum. 
We used the functional form of CXB reported by \citet{gruber99}
(see equation~1 in their paper).

We use the RMF for a point source when we fit the afterglow 
spectrum. However, the CXB is the extended emission, so we have to 
convert the CXB spectral parameters into an equivalent parameters 
for the RMF of point source. To do so, we simulated an expected CXB 
spectrum with the flat field RMF. After that, we fitted the simulated 
CXB spectrum with the point source RMF, and obtained the functional 
form as
\begin{eqnarray}
\frac{dN}{dE} = 8.134 \times 10^{-4} 
\Bigl( \frac{E}{1~{\rm keV}} \Bigr)^{-1.29}
\exp \Bigl( - \frac{E}{41.13~{\rm keV}} \Bigr)
~{\rm photons~keV^{-1}cm^{-2}sec^{-1}}.
\end{eqnarray}
at 10--40~keV range. We used this function as the CXB spectrum.

\begin{figure}
\rotatebox{270}{\includegraphics[width=60mm]{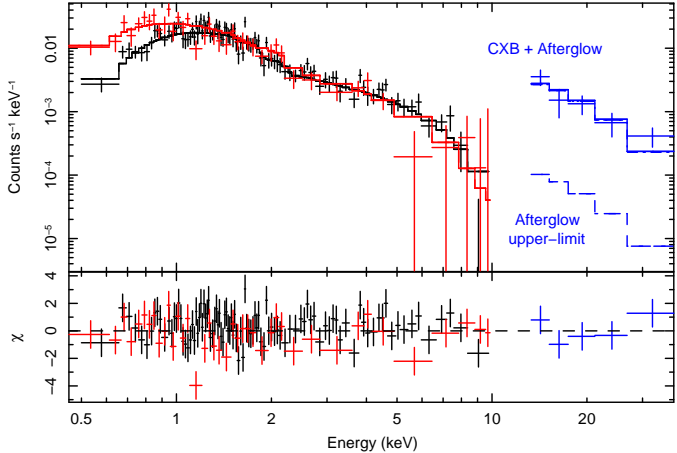}}
\caption{The spectrum of X-ray afterglow of GRB~060904A observed by
the {\it Suzaku}/XIS and HXD-PIN. The black and red colors indicate 
a summed spectrum of front-illuminated chips and one of 
back-illuminated chip, respectively. The blue spectrum includes 
the CXB component, and the upper-limit flux is described by 
the dotted line. Each spectrum includes its own detector responses.}
\label{fig:suzaku-spec}
\end{figure}

In figure \ref{fig:suzaku-spec}, we show the averaged spectrum 
observed with the {\it Suzaku}/XIS and HXD-PIN. 
Adopting the absorbed power-law model, we determined the photon 
index of $\Gamma = 2.00 \pm 0.09$ and the column density of 
$N_{\rm H}^{ext} < 2.30 \times 10^{22}~{\rm cm^{-2}}$ including
systematic error caused by the contamination effect.
The observed flux in 2--10~keV band is
$F_{2-10} = (3.69 \pm 0.31) \times 10^{-13}~{\rm erg~cm^{-2}s^{-1}}$.
Assuming the same photon index of $\Gamma = 2.00$ at the HXD-PIN
energy band, we set the upper-limit flux in 10--40~keV band
as $F_{10-40} < 1.15 \times 10^{-11}~{\rm erg~cm^{-2}s^{-1}}$. 

\section{Advanced Spectral Analyses}
\begin{figure}
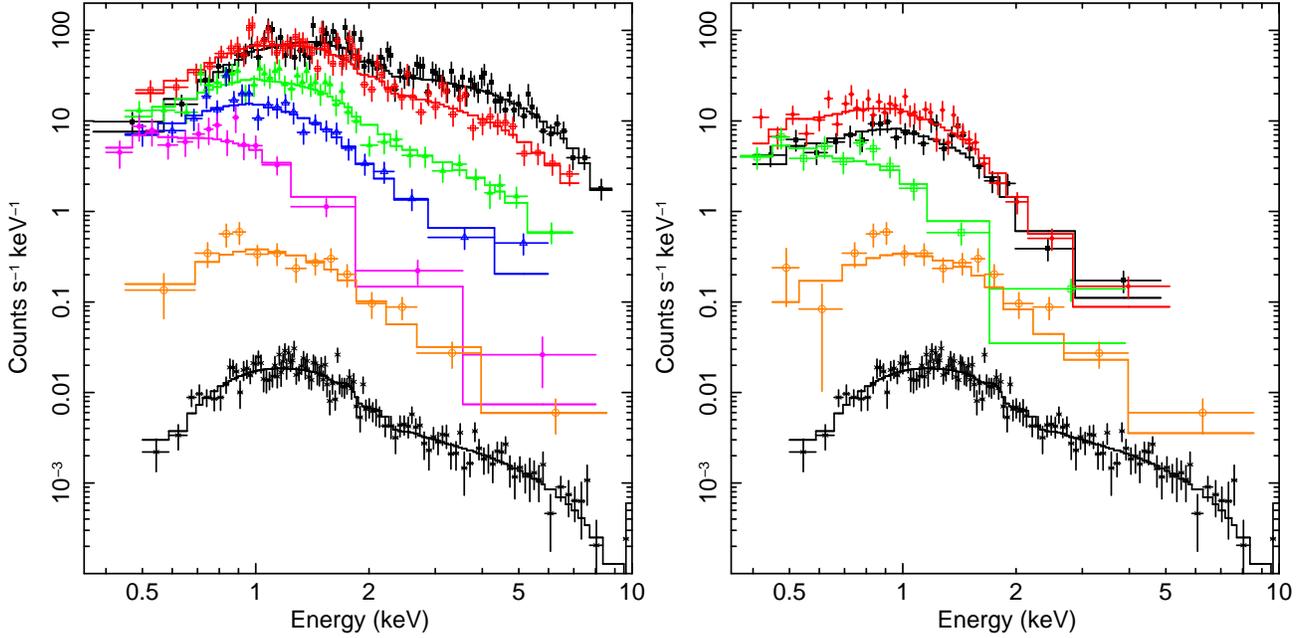

\rotatebox{270}{\includegraphics[width=84mm]{fig6a.ps}}
\rotatebox{270}{\includegraphics[width=84mm]{fig6b.ps}}
\caption{The spectral evolution of early X-ray afterglow of GRB~060904A
with the best fit power-law model listed in table~\ref{table:single-po}.
The left panel shows 7 time-resolved spectra 
(No.1:black, 3:red, 7:green, 9:blue, 12:magenta, 21:orange 
see figure~\ref{fig:xrt-lc}) observed by XRT as well as the late time 
X-ray afterglow observed by {\it Suzaku}/XIS (bottom black).
The right panel shows same as the left one for the X-ray flare around 
$\sim 300~{\rm sec}$. Each color corresponds to the spectrum of 
different time intervals (13:black, 14:red, 16:green and 21:orange)
with the XIS spectrum (bottom black).
For clear comparison with the spectrum of shallow decay phase,
we also inserted the No.21 and late afterglow spectra on these figures.}
\label{fig:spec-evolution}
\end{figure}

In figure~\ref{fig:spec-evolution}, we show several 
representative spectra in different time intervals on the same figure. 
We can clearly recognize the strong spectral softening. 
The flux in the harder X-ray band shows a rapid decline while below 
1.0~keV there is hardly decay at all. This trend was clearly observed 
during the decay phase of the first X-ray flare as shown in 
figure~\ref{fig:spec-evolution} (right). 
If we assume the standard synchrotron radiation by the 
accelerated electrons with a power-law energy distribution 
($N(\gamma_{e}) \propto \gamma_e^{-p}$), the energy index of 
photon spectrum can be described as $\nu^{-p/2}$ in 
the fast cooling regime. For this case, the photon 
index of $\Gamma = 5.3$ is equivalent to an electron energy 
distribution of $\gamma_e^{-8.6}$. This ultra steep index rules
out this efficient acceleration any more.
Although the simple power-law model is acceptable as shown in 
table~\ref{table:single-po}, the ultra soft spectrum deviates
from the standard synchrotron scenario. Some advanced spectral 
models may be required. In this section, we give a possible model 
to explain the observed softening.

The spectra of prompt emissions can be well described by the empirical 
Band function \citep{band93}, which is composed of an exponentially 
connected broken power-law. The accurately determined 
parameters of photon indices for the low and high energy band 
are $\alpha \sim -1$ and $\beta \sim -2.25$, respectively 
\citep{preece00, kaneko06}. When we notice the fitting results
by the single power-law model in table~\ref{table:single-po},
we can recognize that the break energy (peak energy) passes through
the XRT energy window during the time interval 1--6 because 
the photon indices changes from $\Gamma = 1.5$ to softer value than 
$\Gamma = 2$. However, it is not possible to explain the very steep 
photon index, such as $\Gamma = 5.3$, found in GRB~060904A spectra.

\begin{figure}
\rotatebox{0}{\includegraphics[width=90mm]{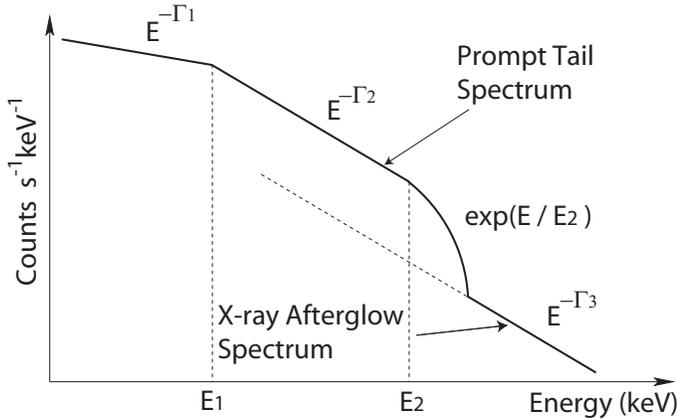}}
\caption{A schematic view of the broken power-law and the exponential
cutoff (BPEC) model.}
\label{fig:model-function}
\end{figure}

Next, we performed an advanced spectral analysis using 
a broken power-law with exponential cutoff (BPEC) model. 
Here we used broken power-laws instead of the Band function 
because the break energy on the Band function can be determined 
only when $10 \le E_0 \le 10^4~{\rm keV}$ in the standard analysis 
package ``XSPEC'' while we hope to measure the break energy in
the XRT band. In the BPEC model, we denote several parameters 
as following (see also figure~\ref{fig:model-function}):
\begin{enumerate}
\item $E^{-\Gamma_1}$ : $\Gamma_1 \sim 1.5$ between $\nu_c < \nu < \nu_m$
for the fast cooling case. Here $\nu_c$ and $\nu_m$ are the frequency 
determined by the cooling time and the minimum electron energy, 
respectively. Hereafter we denote the break energy as $E_1 \equiv h \nu_m$ 
which corresponds to the peak energy in the Band function.
\item $E^{-\Gamma_2}$ : $\Gamma_2 \sim 2.25$ at the range of $\nu > \nu_m$ 
for the power-law index of electron energy distribution of $p=2.5$ 
(fast cooling case). 
\item $\exp(-E/E_2)$ : We introduce a cutoff component above the higher
energy ends to describe the spectra with very steep photon indices. 
Several possible physical interpretations will be discussed in 
the following section.
\item $E^{-\Gamma_3}$ : an additional power-law component is required
to describe a steady hard spectrum of the shallow decay phase.
\end{enumerate}

We succeeded in fitting 22 datasets of XRT spectra with 
the BPEC model. The best fit results are summarized in 
table~\ref{table:bpec}. The blank spaces mean that the BPEC model is 
out of energy range covered by the XRT. 
We found that $E_1$ (equivalent to the peak energy) 
passed through the XRT energy band during the time interval of No.3--5. 
Next, the spectra can be described by the single power-law with 
the photon index of $\Gamma_2 \sim 2.25$. After that, the exponential 
cutoff energy ($E_{2}$) got through the XRT energy window during the
time interval of No.9--16. Finally, the variable prompt tail emission
left the energy range below 0.5~keV, and only the X-ray afterglow
in shallow decay phase was observed.
The spectra obtained by the {\it Suzaku} observation can be well 
fitted by the single power-law with $\Gamma_3 \sim 2$ which have
been already listed in the table~\ref{table:single-po}. 

We also adopted the BPEC model to the BAT data obtained before 
the start time of XRT observation. We used the time interval 
of $55 < t - t_{trigger} < 66$~s which corresponds to the tail 
emission phase of the most intense peak before the XRT 
observation start time. We succeeded in measuring the break energy 
of $E_{1} = 112.2^{+41.6}_{-15.7}$ when we set 
$\Gamma_{2} = 2.25$(fix). This result strongly supports that 
the break energy $E_{1}$ continuously passed through from 
the BAT to the XRT energy ranges.

\begin{table}
\caption{Fitting results of the broken power-law with exponential cutoff model.}\label{table:bpec}
\begin{center}
\begin{tabular}{ccccccccc}
\hline
\hline
No. & Data & $\Gamma_{1}$  & $E_{1}$ & $\Gamma_{2}$ & $E_{2}$ & $\Gamma_3$ & $N_{\rm H}^{ext}$ & $\chi_\nu^2$/dof  \\
 & & & (keV) & & (keV) & afterglow & ($10^{22}~{\rm cm^{-2}}$) & \\
\hline
  & Konus & $1.00^{+0.23}_{-0.17}$ & $163 \pm 31$ & $2.57^{+0.37}_{-1.00}$ & ..... & ..... & ..... & 0.61/61\\
  & BAT   & $1.39^{+0.06}_{-0.06}$ & $112.2^{+41.6}_{-15.7}$& 2.25(fix) & ..... & ..... &  ..... & 0.68/55\\
1 & XRT/BAT & $1.48^{+0.04}_{-0.05}$ & $62.4^{+16.0}_{-17.5}$ & 2.25(fix) & ..... & ..... &  $4.77^{+0.94}_{-0.61}$ & 0.96/141\\
2 & XRT/BAT & $1.54^{+0.06}_{-0.08}$ & $41.4^{+19.3}_{-16.0}$ & 2.25(fix) & ..... & ..... &  $3.16^{+0.77}_{-0.64}$ & 1.16/111\\
3 & XRT/BAT & 1.50(fix) & $8.62^{+3.27}_{-2.85}$ & 2.25(fix) & ..... & ..... &  $1.61^{+0.37}_{-0.31}$ & 1.06/101\\
4 & XRT/BAT & 1.50(fix) & $3.14^{+0.97}_{-0.71}$ & 2.25(fix) & ..... & ..... &  $1.83^{+0.57}_{-0.70}$ & 0.96/69\\
5 & XRT & 1.50(fix) & $2.28^{+0.86}_{-0.42}$ & 2.25(fix) & ..... & ..... &  $1.75^{+0.50}_{-0.42}$ & 1.18/34\\
6 & XRT & 1.50(fix) & $< 1.50$ & $2.18^{+0.23}_{-0.23}$ & ..... & ..... &  $2.22^{+1.40}_{-0.93}$ & 0.76/26\\
7 & XRT & 1.50(fix) & $< 1.42$ & $2.17^{+0.21}_{-0.19}$ & ..... & ..... &  $1.37^{+1.01}_{-0.52}$ & 1.14/38 \\
8 & XRT & ..... & ..... & $2.53^{+0.28}_{-0.25}$ & $> 5.78$  & ..... &  $2.24^{+0.78}_{-0.66}$ & 1.21/25 \\
9 & XRT & ..... & ..... & 2.25(fix) & $3.59^{+3.80}_{-1.41}$ & ..... &  $1.81^{+0.57}_{-0.47}$ & 1.16/21 \\
10 &XRT & ..... & ..... & 2.25(fix) & $1.69^{+0.82}_{-0.49}$ & ..... &  $2.28^{+0.91}_{-0.75}$ & 0.83/18 \\
11 &XRT & ..... & ..... & 2.25(fix) & $1.52^{+0.72}_{-0.46}$ & ..... &  $1.53^{+0.55}_{-0.46}$ & 0.73/19 \\
12 &XRT & ..... & ..... & 2.25(fix) & $0.52^{+0.32}_{-0.17}$ & ..... &  $1.46^{+0.49}_{-0.77}$ & 0.97/14 \\
\hline
13 &XRT & ..... & ..... & 2.25(fix) & $1.51^{+0.64}_{-0.41}$ & ..... &  $1.83^{+0.62}_{-0.48}$ & 1.16/22 \\
14 &XRT & ..... & ..... & 2.25(fix) & $1.09^{+0.30}_{-0.23}$ & ..... &  $1.70^{+0.41}_{-0.35}$ & 1.12/34\\
15 &XRT & ..... & ..... & 2.25(fix) & $0.42^{+0.11}_{-0.08}$ & ..... &  $2.13^{+0.31}_{-0.28}$ & 0.84/21 \\
16 &XRT & ..... & ..... & 2.25(fix) & $0.25^{+0.16}_{-0.10}$ & 2.25(fix)&  $1.66^{+0.56}_{-0.52}$ & 1.27/8 \\
17 &XRT & ..... & ..... & ..... & ..... & $2.25^{+0.41}_{-0.23}$ &  $< 0.94$ & 1.19/10 \\
\hline
18 &XRT & ..... & ..... & 2.25(fix) & $2.23^{+2.17}_{-0.89}$ & &  $1.53^{+0.65}_{-0.50}$ & 0.91/17 \\
19 &XRT & ..... & ..... & ..... & ..... & $2.42^{+0.35}_{-0.31}$ &  $1.08^{+0.69}_{-0.54}$ & 1.00/17 \\
20 &XRT & ..... & ..... & ..... & ..... & $2.25^{+0.73}_{-0.58}$ &  $2.76^{+3.21}_{-2.21}$ & 0.40/6 \\
21 &XRT & ..... & ..... & ..... & ..... & $2.55^{+0.50}_{-0.40}$ &  $2.71^{+1.86}_{-1.33}$ & 1.03/12 \\
22 &XRT & ..... & ..... & ..... & ..... & $2.81^{+0.61}_{-0.46}$ &  $4.08^{+2.19}_{-1.53}$ & 0.92/13 \\
\hline
\end{tabular}
\end{center}
\end{table}

\section{Afterglow Component in the Prompt Tail and the X-ray Flare}
In the previous section, we divided the {\it Swift} and {\it Suzaku} 
data into short time intervals to investigate the detailed temporal 
history of spectral evolutions. The spectra of prompt tail emission 
significantly become softer, and we detect passage of the typical 
break energy through the {\it Swift}/XRT window. 
However, in these analyses, it is difficult to distinguish 
the X-ray afterglow from the prompt tail and/or the X-ray flare 
because the photon statistics at the high energy band become poor. 
Therefore, to improve this fact, we combined several spectra with 
the photon indices softer than $\Gamma > 4$ for the purpose of 
investigating the spectral shape up to 10~keV. 

\begin{figure}
\rotatebox{270}{\includegraphics[width=80mm]{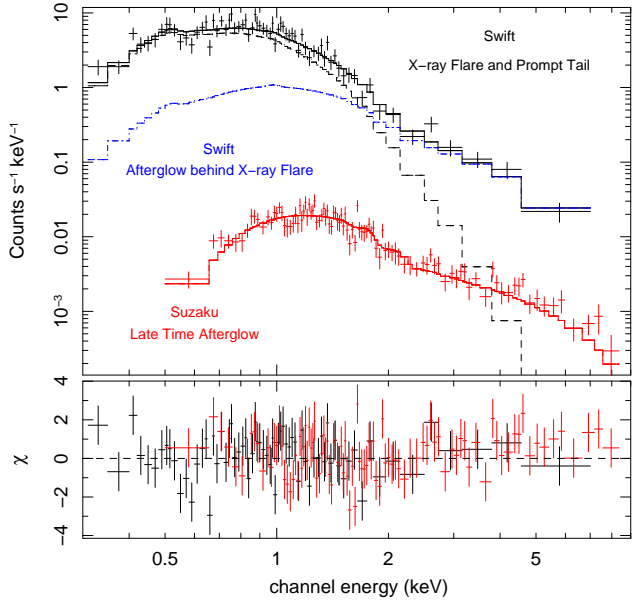}}
\caption{A summed spectrum during the time interval with the photon 
index $\Gamma > 4$ (No.12+15+16). The spectrum breaks around 2~keV, 
and the photon index changes harder toward the higher energy band. 
This spectrum can be fitted with the cutoff power-law plus the
single power-law model representing the prompt tail and 
the behind X-ray afterglow, respectively.}
\label{fig:afterglow-flare}
\end{figure}

In figure~\ref{fig:afterglow-flare}, we show the X-ray spectrum at 
the merged time intervals of 12 + 15 + 16 (black), and the average 
spectrum of late time afterglow observed by {\it Suzaku} (red). 
For this merged spectrum, the BPEC model was rejected with the reduced
chi-square value of $\chi^{2}_{\nu}/\rm{(dof)} = 1.89/(39)$.
This is because the spectrum shows a clear hardening break around 2~keV
and there are large discrepancies between the data and the BPEC model
function above 2~keV. To represent the observed harder spectrum beyond
2~keV, we added a single power-law model of $\propto E^{- \Gamma_3}$
to the BPEC model as described in the previous section.
Then the fitting result is significantly improved as
$\chi^{2}_{\nu}/\rm{(dof)} = 1.15/(38)$. The best fit parameters are
$N_{\rm H}^{ext} = 2.24 \pm 0.25$, $E_2 = 0.29 \pm 0.06$ and 
$\Gamma_3 = 1.87 \pm 0.61$ with the fixed index of $\Gamma_2 = 2.25$
(the other parameters, $\Gamma_1$ and $E_1$, are out of XRT range). 

Therefore we can naturally recognize this spectrum includes 
two emission components. The BPEC model obviously represents the
prompt tail spectrum, and the additional single power-law may represent
the X-ray afterglow component because the photon index ($\Gamma_3$) 
is consistent with one of the shallow decay phase and the late time
afterglow observed with {\it Suzaku} while the uncertainty is quite
large. We succeeded in distinguishing the X-ray afterglow component 
from the prompt tail emission. Although the prompt tail emission shows 
significant spectral evolution as previously shown,
the spectral slope above 2~keV is consistent with the typical 
absorbed power-law with the photon index of $\Gamma \sim 2$.
This fact indicates that the emitter of prompt emission is due to 
completely different dynamics from the X-ray afterglow. 
We can conclude the emission sites of two distinct phenomena 
obviously differ from each other.

\section{Discussions}
We found significant spectral softening during the prompt 
tail emission of GRB~060904A. Moreover, we found similar 
spectral softening in the X-ray flare. The photon indices 
evolve from $\Gamma = 1.5$ to $\Gamma = 5.3$, which is one of 
the steepest spectra ever reported. We think there are 
two noticeable points to be discussed. 
One is a possible origin of the ultra steep spectrum and/or 
the exponential cutoff. And another is the time profile of 
the spectral break energy. 

\subsection{Possibility of Ultra Soft Spectra}

The photon indices in the early X-ray afterglow of GRB~060904A 
evolves toward $\Gamma = 5.3$. This is one of the steepest case 
ever observed.
\citet{bbzhang06} performed simultaneous studies for 17 prompt
tail emissions observed by {\it Swift}/XRT.
According to their results, 10~GRBs have significant hard-to-soft
trend and the others show no spectral evolution.
In this paper, we investigated two spectral models to describe 
the spectral softening -- the simple power-law model and 
the broken power-law with exponential cutoff (BPEC) model.
Although we obtained acceptable results for both models, 
we suggest the BPEC model is better to explain the observed 
spectra as following reasons. 

General X-ray afterglows are well fitted by a simple power-law
model with a photon index of $\Gamma \sim 2$. The origin of 
this non-thermal spectrum is thought to be synchrotron radiation
via accelerated electrons with the power-law energy distribution 
($N(\gamma_e) \propto \gamma_e^{-p}$). Then we expect 
the power-law spectrum with the energy index of $E^{-p/2}$
for the fast cooling regime. 
As previously mentioned, the energy distribution of seed 
electrons with $\gamma_e^{-8.6}$ is required to explain 
the observed photon index of $\Gamma = 5.3$ by the standard 
synchrotron radiation. This is a really steep electron spectrum, 
and we may be able to recognize the existence of high energy 
limit of Fermi-type acceleration if we trust the synchrotron 
scenario. In this case the photon spectrum should be modified. 
The spectral shape changes the exponential function above 
the synchrotron frequency corresponding to the maximum energy 
electrons. The highest end of the BPEC model represents 
this situation, and it can be well adopted to the observed 
spectra for GRB~060904A.

\begin{figure}
\rotatebox{270}{\includegraphics[width=65mm]{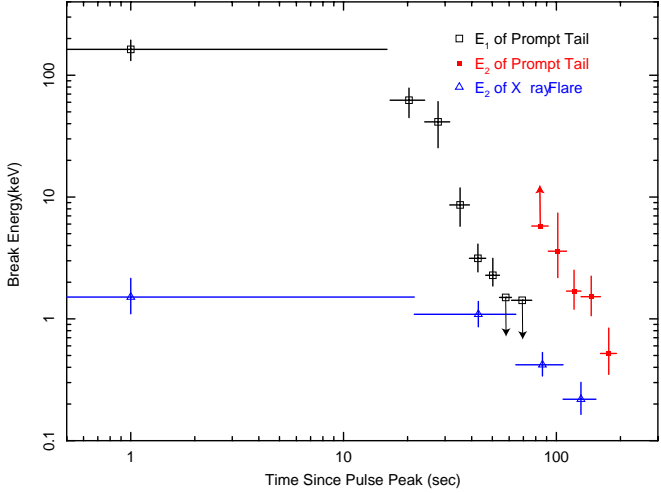}}
\caption{Time histories of spectral break energies. 
The open and filled squares mean the break energy $E_{1}$ and $E_{2}$ 
of the prompt tail emission measured from the most intense
pulse at $t-t_{trigger} = 55~{\rm sec}$. The open triangles mean 
the $E_{2}$ history of the X-ray flare measured from 
$t-t_{trigger} = 268.1$. The decay profiles of $E_{1}$ and $E_{2}$ 
in prompt tail are very similar, and the best fit power-law model 
is $t^{-3.9 \pm 0.4}$ and $t^{-3.2 \pm 1.4}$, respectively.}
\label{fig:break-profile}
\end{figure}

In figure~\ref{fig:break-profile}, we show the time history of 
the spectral break energies. The open and filled squares indicate 
the $E_{1}$ and $E_{2}$ profiles of the prompt tail emission, 
measured from the most intense peak at $t-t_{trigger} = 55~{\rm s}$, 
respectively. The first point of open square is represented by 
the peak energy of $E_{p} = 163 \pm 31$ measured from the average 
spectrum observed with {\it Konus-Wind} \citep{golenetskii06}. 
The open triangles are one of $E_{2}$ in the X-ray flare measured 
from its peak at $t-t_{trigger} = 268.1~{\rm s}$. 
When we adopt a power-law model for each time profile, we obtained 
\begin{eqnarray}
\label{eq:break-energy}
E_{break} \propto
\left\{
\begin{array}{ll}
t^{-3.9 \pm 0.4} & \mbox{for $E_{1}$ of prompt tail}\\
t^{-3.2 \pm 1.4} & \mbox{for $E_{2}$ of prompt tail}\\
t^{-1.7 \pm 0.5} & \mbox{for $E_{2}$ of X-ray flare}.
\end{array}
\right.
\end{eqnarray}

According to the time history of break energies shown in 
figure~\ref{fig:break-profile}, we can recognize that 
the time profiles of both $E_{1}$ and $E_{2}$ in prompt tail emission 
are very similar to each other. 
This fact may lead us to consider the situation that 
the spectral shape is likely to be stable as the BPEC model 
in the rest frame of emitter, and the typical break energies 
cross toward the low energy band caused by some physical mechanism.
When we extrapolate the cutoff energy ($E_{2}$) toward the
brightest time of main emission at $t-t_{trigger} = 55~{\rm sec}$, 
we estimate $E_{2} \sim 1~{\rm MeV}$ which is only one order
of magnitude larger than the peak energy of $E_{p} = 163 \pm 31$ 
reported by \citet{golenetskii06}. 

\subsection{Time Profile of Spectral Break Energy}
Let us assume an instantaneous emission from the relativistic 
shell, then a delayed emission caused by a geometrical curvature 
of emitting material is expected. This effect is widely called as 
the curvature effect. The relativistic boosting of delayed 
emission is much smaller than one of the on-axis emission. 
Therefore, the characteristic frequencies of prompt emission, 
such as $E_{1}$ and $E_{2}$ in this paper, 
become softer as time goes by.

When we define a point where the radial velocity of expanding 
spherical shell is parallel to the line of sight to be 
$\theta = 0$, a Doppler factor ($\delta$) toward the observer 
at a high latitude angle $\theta$ is given by
\begin{eqnarray}
\delta = \frac{1}{\gamma (1-\beta \cos \theta)}.
\end{eqnarray}
Here $\gamma$ and $\beta = v/c$ are the Lorentz factor and 
the velocity of emitting shell, respectively.
The difference of photon arrival time caused by 
the geometrical curvature can be described as
\begin{eqnarray}
t - t_{0} = \frac{R_{0} (1 - \cos \theta)}{c},
\end{eqnarray}
where $R_{0}$ is the radius of emitting material, 
$t_{0}$ is the arrival time of photon emitted at $\theta = 0$, 
and $c$ is the light velocity. 
For extremely relativistic case, we can adopt 
$\beta \sim 1 - 1/2\gamma^2$ and denote the Doppler factor
as a function of $t$;
\begin{eqnarray}
\delta(t) \sim \frac{2 \gamma}{1 + (t - t_{0})/\tau},~~~
\tau \equiv \frac{R_0}{2 \gamma^2 c}.
\end{eqnarray}
Then the spectral break energies are expected to have 
the time dependence of
\begin{eqnarray}
E_{break}(t) = \frac{E_{break,0}}{1 + (t - t_{0})/\tau}.
\end{eqnarray}
Here, the time constant $\tau$ means an angular spreading time 
and it is equivalent to about 1~second for the typical parameters
of $R_{0} \sim 10^{15}~{\rm cm}$ and $\gamma \sim 100$. 
Therefore the spectral evolution during the prompt tail at 
$t - t_{0} \gg 1~{\rm sec}$ can be approximately described 
as $\propto (t-t_{0})^{-1}$ if the curvature effect controls 
the spectral softening. However, we found the time history of 
characteristic energy $E_{1}$ and $E_{2}$ as 
equation~\ref{eq:break-energy}, and they are inconsistent with 
the above discussions. 

Let us consider the curvature effect from a different point 
of view. The spherical curvature effect predicts the relation 
of $\alpha_{x} = \beta_{x} + 2$ between the temporal index 
$\alpha_{x}$ and the spectral energy index $\beta_{x}$ 
($=\Gamma - 1$) \citep{kumar00}. 
As shown in figure~\ref{fig:single-power}, the temporal index is 
rather stable as $\alpha_{x} = 6.0 \pm 0.2$ during the prompt tail 
while the spectral index evolves from $\Gamma = 1.5$ to 
$\Gamma = 5.3$ for the case of single power-law fitting. 
This observational fact is not satisfied with the above relation. 
For the fitting based on the BPEC model, it is difficult to 
argue the consistency or discrepancy of the 
$\alpha_{x} = \beta_{x} + 2$ relation, because this relation is 
valid only in the case where no spectral break energy crosses 
the observational band. Even if we assume the photon indices 
of pre- and post-break as $\Gamma_1 = 1.5$ and $\Gamma_2 = 2.25$, 
either cases can not adopt the $\alpha_{x} = \beta_{x} + 2$ 
relation. Therefore, we conclude it is impossible to explain the
observed spectral and temporal properties by only 
the spherical curvature effect. We may have to consider 
more complicated system such as the structured jet and 
the multiple jets rather than the spherical uniform jet
as well as hydrodynamics.\\

The spectra of prompt emissions are widely believed as
the Band function \citep{band93} with the low and the high 
energy photon indices of $\sim -1$ and $\sim -2.25$, 
respectively. For the bright GRBs, e.g. GRB~990123, 
the non-thermal emissions is really extending over 
$\sim 10~{\rm MeV}$ \citep{briggs99}. 
\citet{dingus01} reported an average spectrum of 4 GRBs
detected by {\it CGRO}/EGRET. The spectrum really achieves
10~GeV with the photon index of $\Gamma = 1.95 \pm 0.25$.
This is consistent with an extension of the electron
synchrotron component.
On the other hand, \citet{preece00} reported the spectral 
properties for the brightest 156 GRBs detected 
by {\it CGRO}/BATSE. In their results, about 10~\% of 
treated spectra show the high energy indices of $\beta < -4$.
\citet{kaneko06} independently confirmed the same property. 
They categorized these events showing steep spectra as 
a non-high energy portion.

In this paper, we showed the very steep spectra in the early
X-ray afterglow, and argued the possibility of exponential
cutoff at the highest part of X-ray spectrum.
If this GRB~060904A and the events with spectral evolution 
reported by \citet{bbzhang06} belong to the non-high energy 
population defined by \citet{preece00}, the fraction of 
inefficient acceleration cases of all GRBs may be larger 
than we have expected, and reaches about 50~\%.
Then, several physical phenomena may have to be reconsidered. 
This spectral evolution will be an important key to investigate
the acceleration mechanism in the relativistic blast wave.
We encourage the future observation with the {\it Swift}/BAT 
and XRT for early X-ray afterglows of very bright GRBs.

\section{Acknowledgment}
We thank R. Yamazaki, K. Ioka, T. Sakamoto and G. Sato 
for the useful discussions about the spectral evolution 
based on the curvature effect. We also thank 
This research was supported by Grant-in-Aid for 
Scientific Research of the Japanese Ministry of Education, 
Culture, Sports, Science and Technology, No.18684007 (DY).
YEN is supported by the JSPS Research Fellowships for 
Young Scientists.

\end{document}